\documentclass[conference, 10pt]{IEEEtran}
\ifCLASSINFOpdf
\else
\fi
\usepackage{graphicx}
\usepackage{color}
\usepackage{placeins}
\usepackage{float}
\usepackage{hyperref}
\usepackage{tabularx,colortbl}
\usepackage{subfigure}

\usepackage{graphicx,epsfig,epstopdf}
\usepackage{amsmath}

\usepackage{array}

\usepackage{multirow}
\usepackage[utf8]{inputenc}

\begin{document}
%
\title{Graphene-based Perfect Absorbers: Systematic Design and High Tunability}




\author{\IEEEauthorblockN{Xuchen Wang$^*$,
Sergei A. Tretyakov}
\IEEEauthorblockA{Department of Electronics and Nanoengineering\\
 Aalto University, P.~O.~Box~15500, FI-00076 Aalto, Finland\\ Email: xuchen.wang@aalto.fi}}


\maketitle

\begin{abstract}
Experimental realization of efficient graphene-based absorbers is a challenging task due to the low carrier mobility in processed graphene.
In this paper, we circumvent this problem by placing uniform graphene sheets on metallic metasurfaces designed for improving the absorption properties of low-mobility graphene. Complete absorption can be achieved for different frequencies with the proper metasurface design. 
In the THz band, we observe strong tunability of the absorption frequencies and magnitudes when modulating graphene Fermi level.
\end{abstract}


%
\IEEEpeerreviewmaketitle

\section{Introduction}
In the literature, most of the theoretical works predict perfect absorption in graphene with the assumptions of high mobility ($\mu>10000$~${\rm cm}^{2}{\rm V}^{-1}{\rm s}^{-1}$) and high Fermi level \cite{thongrattanasiri2012complete}.  However, the typical achievable mobility in experiment is around $\mu=1000$~${\rm cm}^{2}{\rm V}^{-1}{\rm s}^{-1}$ \cite{kim2017electronically}, which makes these theoretical models impractical to be implemented. 
Despite the large efforts devoted to improve the absorption properties of graphene, it has been shown that a monolayer graphene behaves as a poor absorber in almost all the frequencies.  
From the circuit-theory perspective, the main reason for the weak absorption in graphene is the severe impedance mismatch between graphene and its surrounding materials (the sheet impedance of undoped graphene is usually several thousand ohm per square or more).

In this talk, we will present and discuss the use of metasurfaces for reducing the effective impedance of graphene, i.e., for improving the absorption efficiency in graphene. 
With this method, even in very poor quality graphene samples ($\mu=300$~${\rm cm}^{2}{\rm V}^{-1}{\rm s}^{-1}$), the realization of perfect absorption is possible. 
In addition, we will show how the proposed structures are highly tunable for different frequencies or levels of absorption in the THz band. 
Readers can get more detailed information in~\cite{wang2017tunable}.


\section{Theory }\label{section:Theory}

At THz frequencies or below, the surface conductivity of graphene is described by the Drude model, $\sigma_{\rm g}={e^2E_{\rm F}}/{\pi\hbar^2(j\omega+\gamma)}$, where $\omega$ is the angular frequency, $E_{\rm F}$ is the Fermi level,  and $\gamma={e{v_{\rm F}}^2}/({\mu E_{\rm F}})$ is the scattering rate (related to carrier mobility $\mu$ and Fermi level $E_{\rm F}$). 
Using an equivalent circuit model, the graphene sheet can be considered as a complex-impedance sheet, with the sheet impedance $Z_{\rm g}={1}/{\sigma_{\rm g}}=R_{\rm g}+jX_{\rm g}$. 

\begin{figure}[h!]
	\centering
	\subfigure[]{
		\includegraphics[width=0.35\linewidth]{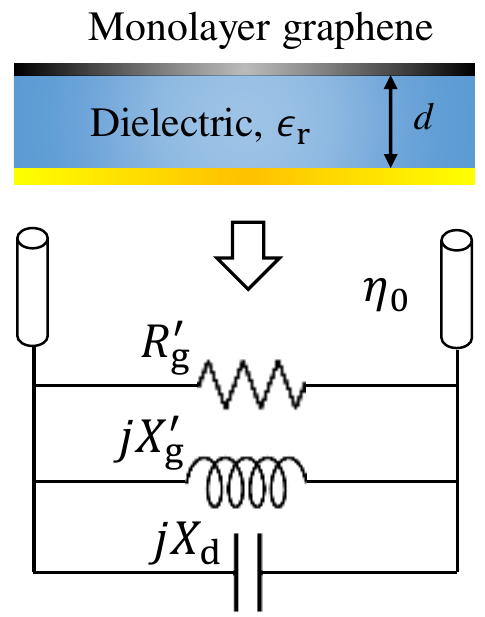}
		\label{fig:Salisbury_circuit}}
	\subfigure[]{
		\includegraphics[width=0.45\linewidth]{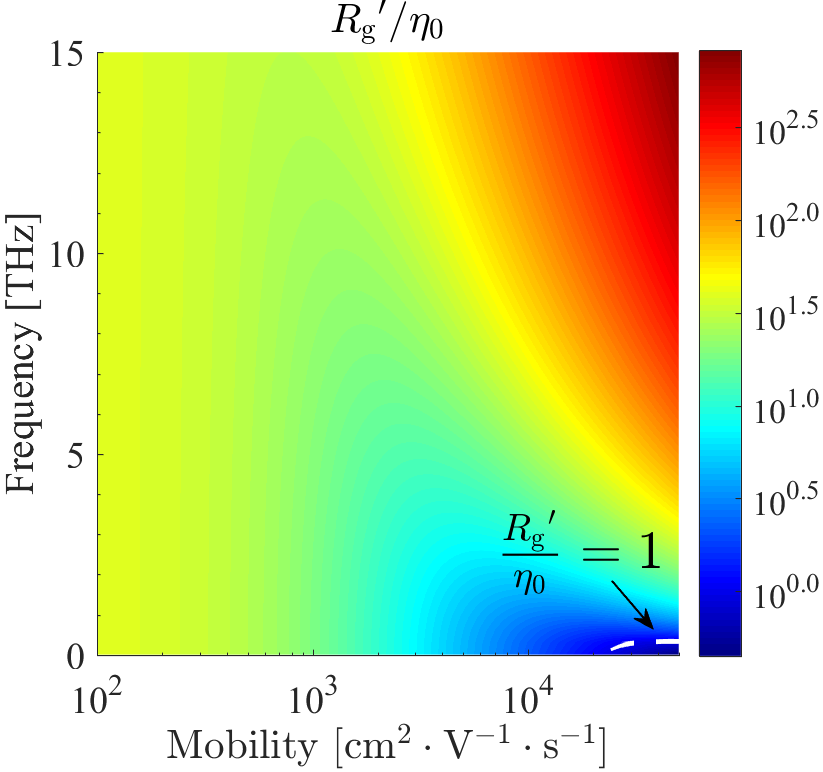}
		\label{fig:Rg_p_EF_0_1}}
	\caption{(a) Structure of graphene Salisbury screen and its equivalent circuit. (b) Normalized shunt resistance of graphene in terms of the carrier mobility and frequency.}\label{salisbury}
\end{figure}

Let us first discuss the absorption characteristics of a graphene Salisbury screen where the graphene layer is supported by a grounded substrate, as depicted in Fig. \ref{fig:Salisbury_circuit}. The equivalent circuit of this structure is described as a parallel $RLC$ resonant circuit. 
Instead of expressing graphene sheet impedance as a series connection of resistance and reactance, here we interpret graphene as a resistor $R_{\rm g}^\prime$ shunt connected with an inductor $jX_{\rm g}^\prime$ ($Z_{\rm g}=R_{\rm g}^\prime\parallel jX_{\rm g}^\prime$), where 
\begin{equation}
R_{\rm g}^\prime=R_{\rm g}+\frac{X_{\rm g}^2}{R_{\rm g}}
\label{shunt_Rg},~\text{and} \quad jX_{\rm g}^\prime=j\left(X_{\rm g}+\frac{R_{\rm g}^2}{X_{\rm g}}\right)
\end{equation}
Since the graphene sheet is inductive in the THz band, the substrate thickness $d$ should be between $\lambda_{\rm d}/4$ and $\lambda_{\rm d}/2$ to ensure its impedance $jX_{\rm d}=j\eta_{\rm d}\tan(k_{\rm d}d)$ is capacitive and thus create a high impedance surface at the resonant frequency.

The use of this equivalent circuit model allows systematic design of perfect absorbers: the parallel resistance controls the absorption efficiency, while the inductance together with the capacitive grounded substrate determine the resonant frequency. 
In order to achieve perfect absorption at a resonant frequency, $R_{\rm g}^\prime$ should be equal to free space impedance $\eta_0$. 
We assume the graphene Fermi level is $E_{\rm F}=0.1$~eV in its natural state (no intentional chemical or electrical doping). 
For this value of $E_{\rm F}$, Fig. \ref{fig:Rg_p_EF_0_1} shows the calculated shunt resistance with respect to frequency and  graphene mobility, where one can see that $R_{\rm g}^\prime$ is huge compared to the free space impedance when $\mu<5000$~${\rm cm}^{2}{\rm V}^{-1}{\rm s}^{-1}$. 
This is the main reason why perfect absorption in a complete graphene layer is quite difficult to obtain in experiment. 

\begin{figure}[h!]
	\centering
	\subfigure[]{
		\includegraphics[width=0.67\linewidth]{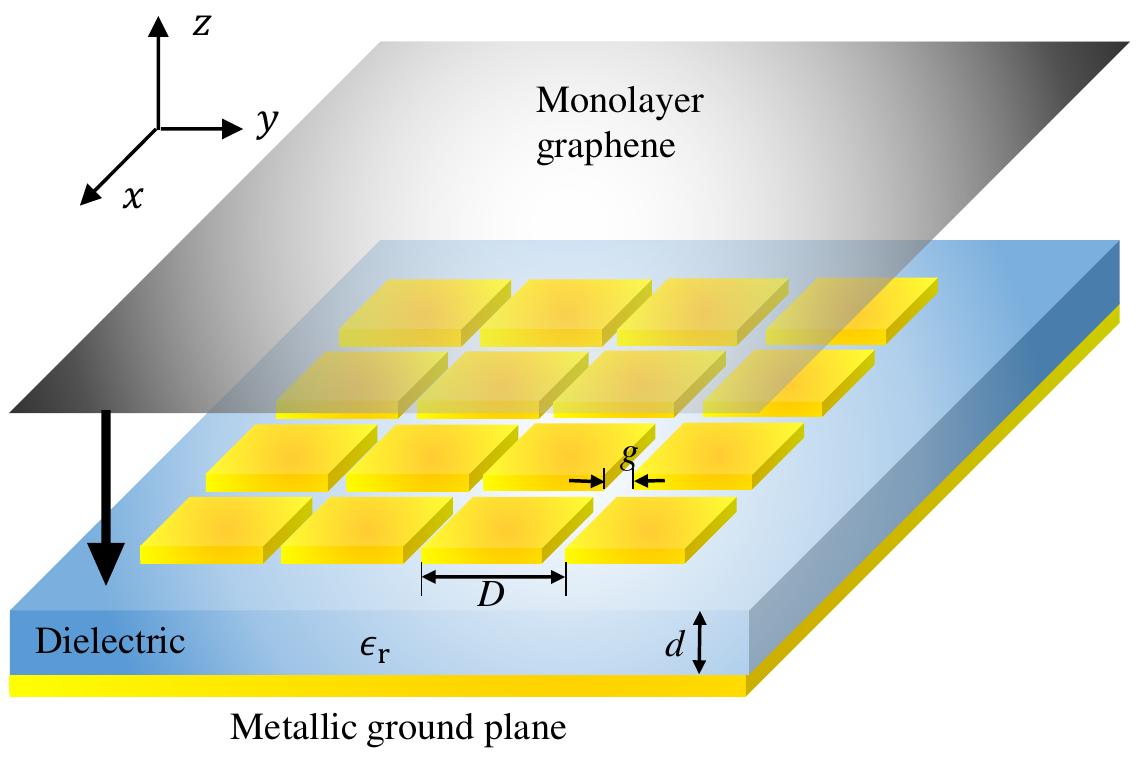}
		\label{fig:proposed_structure}}
	\subfigure[]{
		\includegraphics[width=0.26\linewidth]{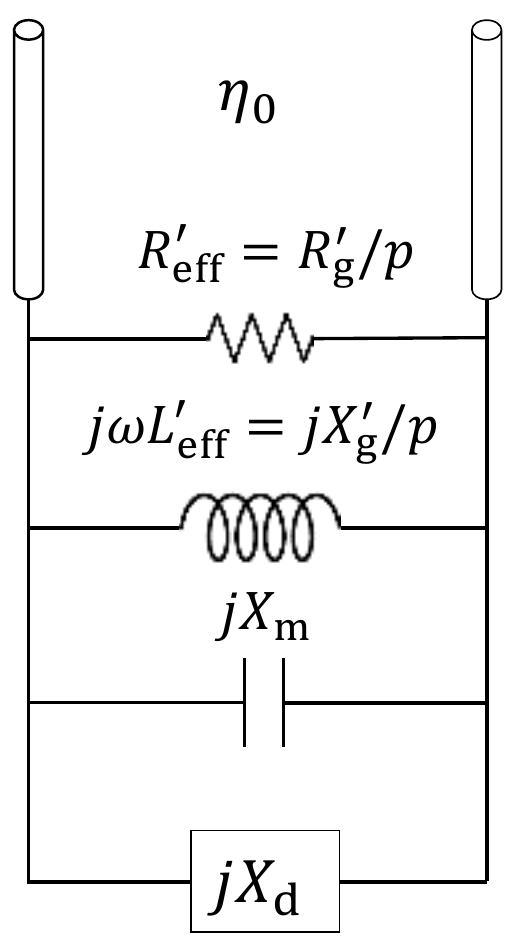}
		\label{fig:proposed_circuit_model}}
	\caption{(a) Schematics of the proposed graphene absorber based on metasurface substrate.  (b) equivalent circuit model of the proposed structure.}\label{proposed_configuration}
	\vspace{-13pt}
\end{figure}

To reduce the shunt resistance of graphene. we replace the substrate with a reflective-type metasurface, as shown in Fig.~\ref{fig:proposed_structure}. The metasurface consists of the grounded substrate and the periodically arranged square patches with the period $D$ and gap size $g$. The graphene sheet is directed transferred onto this metallic metasurface. In the area where graphene and metal are contacted, graphene is actually shorted by the metallic patches. Therefore, the complete graphene sheet is equivalent to be patterned into mesh-type strips with effective impedance $Z_{\rm eff}=Z_{\rm g}/p$, where $p=(D-g)/g$ is called the scaling factor. The grid reactance of periodic patches is expressed as $jX_{\rm m}={1}/{j\omega C_{\rm m}}$ with $C_{\rm m}={\frac{(\epsilon_{\rm r}+1)\epsilon_0D}{\pi}\ln\left(\csc\frac{\pi}{2(p+1)}\right)}$.
In principle, by choosing a suitable scaling factor $p$,  $R_{\rm g}^\prime$ can be matched to the free space impedance $\eta_0$.  By properly choosing the substrate thickness and permittivity or period $D$, the resonant frequency can be adjusted to the expected one.

\section{Electrical tunability}

\begin{figure}[b]
	\centering
	\includegraphics[width=0.73\linewidth]{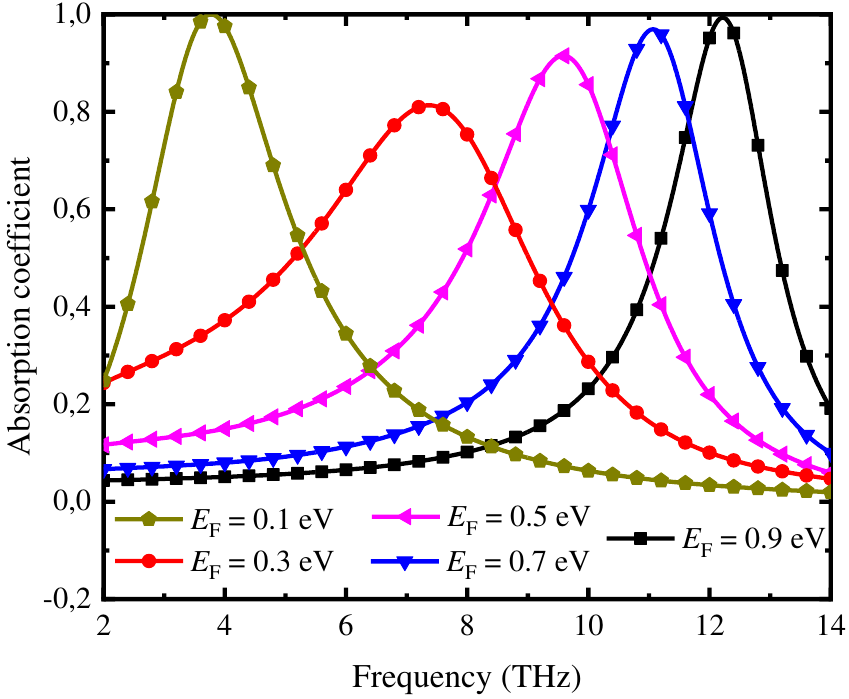}
	\caption{Simulated absorption intensity in terms of the frequency when varying the Fermi level from 0.1~eV to 0.9~eV. The graphene mobility is set to $\mu=1500~{\rm cm}^{2}{\rm V}^{-1}{\rm s}^{-1}$. Here, $D=9.5~\mu$m, $d=4.75~\mu$m, $p=20$ and $\epsilon_{\rm r}=2.8$. }\label{fig:SU8_frequency_tunable}
	\vspace{-20pt}
\end{figure}

In this section, we present two tunable scenarios for absorption frequency and amplitude with different graphene quality. The assumed carrier mobilities in these two cases are typical values in experiment. In the first case, we use graphene with~$\mu=1500$~${\rm cm}^{2}{\rm V}^{-1}{\rm s}^{-1}$. Perfect absorption is designed at 3.8~THz~($E_{\rm F}=0.1$~eV) according to the design rules in Section \ref{section:Theory}. By improving the Fermi level of graphene from 0.1~eV to 0.9~eV, large shift of the peak frequency is observed. The center absorption frequency is tuned from 3.8~THz to 12.3~THz, as shown in Fig. \ref{fig:SU8_frequency_tunable}.

\begin{figure}[h!]
	\centering
	\includegraphics[width=0.73\linewidth]{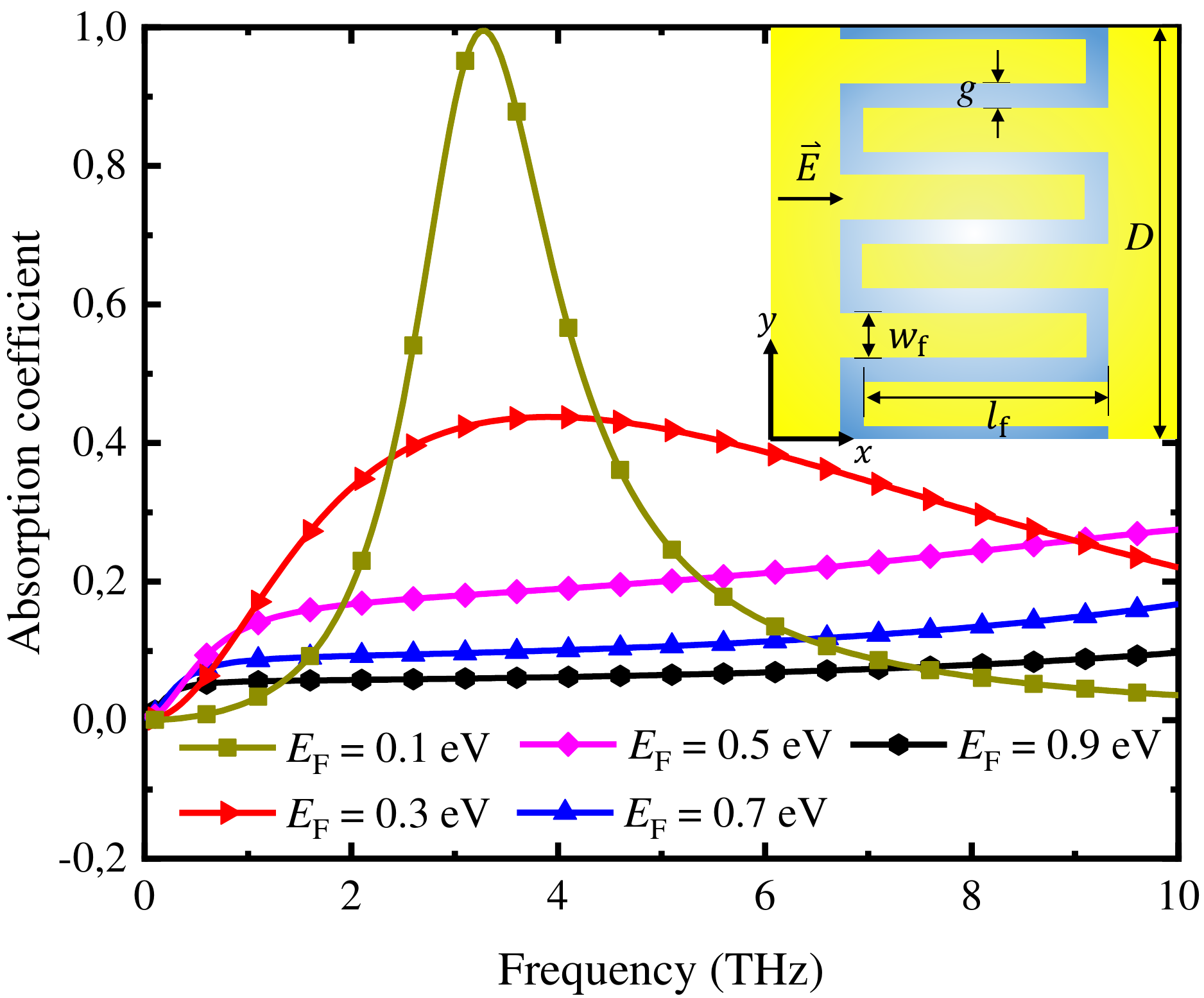}
	\caption{Simulated absorption in terms of the frequency when varying the Fermi level from 0.1~eV to 0.9~eV.  Here, $D=5~\mu$m, $d=4~\mu$m, $g=300$~nm, $l_{\rm f}=3~\mu$m, $w_{\rm f}=533$~nm and $\epsilon_{\rm r}=2.8$. }\label{fig:SU8_intensity_tunable}
	\vspace{-13pt}
\end{figure}

The other scenario is strong tunability of absorption amplitude at a specified frequency. 
We assume very low-quality graphene with $\mu=300$~${\rm cm}^{2}{\rm V}^{-1}{\rm s}^{-1}$. For perfect absorption at 3.3~THz, scaling factor is calculated as $p=91$. If the period $D$ is 5~$\mu$m, the gap size of the patches is as narrow as $g=55$~nm. This nano-scale channel not only brings fabrication challenges, but also results in Fermi-level spinning effect in graphene channels. We replace the straight gap with meandered slot, as shown in the inset of Fig. \ref{fig:SU8_intensity_tunable}. The intersected fingers largely increase the length of graphene channel in one unit thus decreasing the effective resistance of graphene. We can see in Fig. \ref{fig:SU8_intensity_tunable} that the absorption achieves unity with low Fermi levels and almost zero at high Fermi levels, as a switchable absorber.

\section{Conclusions}

In this paper, we use a simple transmission-line model to explain the weak wave absorption in graphene, and propose metallic metasurface substrates to obtain complete and tunable absorption in the THz band. This method is effective even for graphene samples of poor quality.

\section*{Acknowledgment}

This project has received funding from the European Union's Horizon 2020 research and innovation programme-Future Emerging Topics (FETOPEN) under grant agreement No 736876.

\end{document}